\documentclass[12pt]{iopart}
\eqnobysec
\usepackage{iopams} 
\usepackage{epsfig}
\usepackage{amssymb}

\begin{document}
%\pagenumbering{arabic}

\title{Decay of Gaussian correlations in local thermal reservoirs }
\author{Paulina Marian{$^{1,2}$}, Iulia Ghiu{$^1$}
and Tudor A. Marian{$^1$}}
\address{$^1$Centre for Advanced  Quantum Physics,
Department of Physics, University of Bucharest, 
R-077125 Bucharest-M\u{a}gurele, Romania}
\address{$^2$ Department of Physical Chemistry,
University of Bucharest, Boulevard Regina Elisabeta 4-12, R-030018  Bucharest, Romania}
\ead{paulina.marian@g.unibuc.ro}

\begin{abstract}
In this paper we examine the decay of quantum correlations for the 
radiation field in a two-mode squeezed  thermal state in contact 
with local thermal reservoirs. Two measures of the evolving quantum correlations are compared: the entanglement of formation and the quantum discord. We derive analytic expressions of the entanglement-death time 
in two special cases: when the reservoirs for each mode are identical, 
as well as when a single reservoir acts on the first mode only. 
In the latter configuration, we show that all the pure Gaussian 
states lose their entanglement at the same time determined solely
by the field-reservoir coupling. Also investigated is the evolution 
of the Gaussian quantum discord for the same choices of thermal baths. 
We notice that the discord can increase in time above its initial value 
in a special situation, namely, when  it is defined by local measurements on the attenuated mode and the input state is mixed. This enhancement of discord is stronger for zero-temperature reservoirs and increases 
with the input degree of mixing.
\end{abstract}


\maketitle

\section{Introduction}

The extent to which the non-classical properties of one-mode
 field states  survive in the presence of noise
and losses was investigated since the early years of quantum optics \cite{VW,MW}. A routine operation like the transmission of light beams through 
an optical fiber could produce a substantial
degradation of their non-classical properties. As an example, it was found that squeezing properties are altered by admixture with thermal noise and disappear completely for values of thermal mean photon occupancy exceeding the threshold $1/2$ \cite{PT1996}.
  From a more recent quantum-information perspective, a lot of work is concentrated on correlations such as entanglement and discord 
in multi-partite systems \cite{QC1,QC2,QC3}. While correlations 
associated with entanglement \cite{QC1} are defined in connection to
global transformations of bipartite quantum states, the concept of quantum discord arises from local (marginal) actions and measurements performed on
one subsystem \cite{OZ,HV}. Its definition 
 contains an optimization over the set of all
one-party measurements, which in the case of mixed states could be 
a challenging problem.  Note that in the pure-state case, entanglement and discord coincide and therefore they measure the total amount 
of correlations. In the mixed-state case, quantum discord 
is a measure of quantumness whose relation to entanglement 
is not a simple one.  A survey of recent progress and applications of classical and quantum correlations quantified by quantum discord 
and other measures can be found in Refs. \cite{QC2,QC3}.

When states of a two-party quantum system need to interact with a noisy channel, a drastic modification of their quantum correlations  is expected to occur \cite{Yu,Damp}. For instance, 
it was found that  quantum and classical correlations for a system of two qubits evolving in Markovian dephasing channels can display different dynamics \cite{sabrina}.  Quite recently,  the effect of local
noisy channels  on quantum correlations in finite-dimensional quantum systems was investigated \cite{bruss}. It was found that while entanglement does not increase under local channels, other correlations can become larger
when the input state is not pure. In the continuous-variable settings, 
a similar behaviour was recently noticed for the Gaussian discord 
of two-mode mixed states under single-mode Gaussian dissipative channels 
\cite{Ciccarello}.  Local Gaussian thermal and phase-sensitive reservoirs modify the entanglement properties of two-mode Gaussian states, 
as interestingly is pointed out in Refs. \cite{serafini,goyal,souza}. 
Evolutions of more general Gaussian correlations were also investigated 
in Refs. \cite {buono,barbosa,isar,madsen}.

In this work we analyze the decay of quantum correlations of the field
in a two-mode Gaussian state due to the interaction of the modes with  separate thermal baths. We focus on two measures of quantum correlations, namely, the entanglement of formation (EF) 
and the quantum discord,
and evaluate them for a damped two-mode squeezed thermal state (STS) 
\cite{PTH2001,PTH2003}. Our choice of this important particular class 
of Gaussian states is motivated by the recent result \cite{PSBCL} that 
for an STS, the exact discord according to its original definition 
\cite{OZ,HV} is achieved with an optimal measurement which is Gaussian.
Consequently, what was called the Gaussian discord and was derived 
in Refs. \cite{PG,AD} is actually the {\em exact} discord. 
On the other hand, in the interesting paper \cite{OP} it is shown 
that for an STS, a Gaussian character of the discord implies that the EF 
is also Gaussian. Accordingly, the Gaussian EF written explicitly in our paper \cite{PT2008} is equally an exact result. As such, when dealing 
with a dissipative evolution that preserves the STSs, we have the rare 
privilege to fully describe the decay of two types of correlations by analytic means. 

Our paper is structured as follows. In Section 2 we recapitulate
several properties of an STS: the covariance matrix, the Simon separability criterion \cite{Simon}, the entanglement of formation \cite{PT2008} as a measure of inseparability, and the quantum discord as derived 
in Refs. \cite{PG,AD}. In Section 3 the system of interest (field $+$ environment) is specified in order to write and solve the quantum optical master equation. Special attention is then paid to the solution 
for an input STS. Section 4 deals with a damped STS for modes 
coupled with two identical local thermal baths. We here derive 
the time at which the entanglement sudden death occurs and study the evolution of the discord. Section 5 is dedicated to another interesting configuration of the system: only one mode is in contact with a thermal bath. We find that all the pure Gaussian states lose their entanglement 
at the same time which depends on the reservoir only. In the case 
of a mixed input state, the discord defined by local measurements on the attenuated mode is increased above its initial value. Our conclusions are drawn in Sec. 6.

\section{Quantum correlations in a two-mode squeezed thermal state}

In this section we review shortly the above-mentioned examples of Gaussian correlations, with emphasis on those of an STS. We consider two-mode
Gaussian states and let us denote the photon annihilation operators 
of the modes by $\hat a_1$ and $\hat a_2.$  
As shown in Refs. \cite{PTH2001,PTH2003}, an STS is the result 
of the action of a two-mode squeeze operator, 
\begin{eqnarray*}
\hat S_{12}(r,\phi):=\exp{\left[ \,r \left( {\rm e}^{i\phi} 
\hat a^{\dag}_1 \hat a^{\dag}_2-{\rm e}^{-i\phi} \hat a_1 
\hat a_2 \right) \right] },\quad \left( r>0,\;\; 
\phi\in (-\pi,\pi] \right),
\end{eqnarray*} 
on a two-mode thermal state with the mean photon occupancies 
$\bar{n}_1$ and $\bar{n}_2$:
\begin{eqnarray}
 \hat \rho_{ST}=\hat S_{12}(r, \phi)\hat \rho_T (\bar n_1,\bar n_2) 
\hat S^{\dag}_{12}(r, \phi) \label{sts}.
\label{STS}
\end{eqnarray} 
Its covariance matrix (CM) has the following block 
structure \cite{PTH2001,PTH2003}:
\begin{equation}
{\cal V}=\left( \begin{array}{cc}
b_1\, \mathbb{I}_2&{\cal C}\\
{\cal C}&b_2\, \mathbb{I}_2
\end{array} \right),
\quad \left( b_1>\frac{1}{2},\; b_2>\frac{1}{2}\right ).
\label{cv}
\end{equation}
In Eq.\ (\ref{cv}), $\mathbb{I}_2$ denotes the 2$\times $2 identity matrix and ${\cal C}$ is the 2$\times $2 symmetric matrix
\begin{equation}
{\cal C}=c \left( \begin{array}{cc}
\cos{\phi}\quad  \sin{\phi} \\ 
\sin{\phi}\quad -\cos{\phi}
\end{array} \right), \quad (c>0).
\label{mc}
\end{equation}
Recall that the CM of an STS, Eqs.\ (\ref{cv}) and\ (\ref{mc}), 
has the standard-form parameters \cite{PTH2001,PTH2003}: 
\begin{eqnarray}
b_1&=&\left( \bar{n}_1+\frac{1}{2}\right) [\cosh(r)]^2
+\left( \bar{n}_2+\frac{1}{2}\right) [\sinh(r)]^2, \nonumber \\
b_2&=&\left( \bar{n}_1+\frac{1}{2}\right)[\sinh(r)]^2+\left( \bar{n}_2
+\frac{1}{2}\right) [\cosh(r)]^2, \nonumber \\
c&=&(\bar{n}_1+\bar{n}_2+1)\sinh(r) \cosh(r).
\label{par}
\end{eqnarray}

In many applications one can take advantage of a formal definition 
of an STS, as being an undisplaced and unscaled two-mode Gaussian state described by three standard-form parameters: $b_1>\frac{1}{2}, \;
b_2>\frac{1}{2}, \; c>0$. If $b_1\geqq b_2$, then these parameters 
must fulfill the uncertainty inequality
\begin{equation}
\left( b_1+\frac{1}{2}\right) \left( b_2-\frac{1}{2}\right)-c^2 \geqq 0.
\label{state}
\end{equation}
If $b_1 < b_2$, then one has to interchange the parameters 
$b_1$ and $b_2$ in Eq.\ (\ref{state}) \cite{PTH2003}. 
The standard form of its CM is given by Eq.\ (\ref{cv}) with 
the $2\times 2$ matrix ${\cal C}$  written for $\phi=0$, i.e., becoming proportional to the Pauli matrix $\sigma_3:\; {\cal C}= c\,\sigma_3$. Within this formal treatment, 
Eq.\ (\ref{sts}) and its companions, Eqs.\ (\ref{cv})--\ (\ref{par}), 
represent a parametrization of an STS with a clear experimental relevance. 

It is known that, according to Williamson's theorem \cite{Wil}, the CM of a two-mode  Gaussian state can be diagonalized 
by a symplectic transformation. We get thus an important ingredient in describing the state, namely, the symplectic eigenvalues of the CM. 
For an STS they are \cite{PTH2003}:
\begin{eqnarray}
\kappa_{\pm}=\frac{1}{2}\left[ \sqrt{(b_1+b_2)^2-4c^2}\pm (b_1-b_2)\right].
\label{se}
\end{eqnarray}
In the parametrization\ (\ref{par}), we get $ \kappa_{\pm}=\bar n_{1,2}
+\frac{1}{2}$.

It is worth mentioning Simon's separability criterion for two-mode 
Gaussian states \cite{Simon}. It was proven that preservation 
of the non-negativity of the density matrix under partial transposition 
is not only a necessary \cite{Peres}, but also a sufficient condition 
for the separability of two-mode Gaussian states \cite{Simon}. 
Accordingly, a two-mode  Gaussian state is separable when the condition 
$\tilde \kappa_-\geqq \frac{1}{2}$ is met. We have denoted 
by $\tilde \kappa_{\pm}$ the symplectic eigenvalues of the CM 
corresponding to the partial transpose of the density matrix. 
For an STS one finds:
\begin{eqnarray}
\tilde \kappa_{\pm}=&\frac{1}{2}\left[b_1+b_2 \pm\sqrt{(b_1-b_2)^2
+4c^2}\right].
\label{pse}
\end{eqnarray}
A simplified form of the separability condition for an STS 
that we shall use in what follows reads \cite{PTH2001,PTH2003}:
\begin{eqnarray}
\left({b_1}-\frac{1}{2}\right)
\left({b_2}-\frac{1}{2}\right)-{c}^2\geqq 0.
\label{sc}
\end{eqnarray}
Before proceeding, let us note that, apart from the vacuum state, the only undisplaced and unscaled pure two-mode Gaussian states are the squeezed vacuum ones, $|{\psi}_{SV}\rangle\langle {\psi}_{SV}|$, and they belong 
to the set of STSs. The pure-state case is characterized by the identities $b_1=b_2=:b$ and $b^2-c^2=\frac{1}{4}$. Equations\ (\ref{se}) 
and\ (\ref{pse}) give now 
$\kappa_{\pm}=\frac{1}{2},\;\; \tilde \kappa_{\pm}=b\pm c$.

In the following, we shall recall two measures of quantum correlations for an STS, namely, the EF and the quantum discord. 
The EF is defined as an optimization over all the  pure-state 
decompositions  of the given state \cite{Bennett}:
\begin{eqnarray}
E_F(\hat \rho ):=\inf \left[ \sum_kp_k\, E\left( |\psi_k
\rangle \langle \psi_k|\right) \; | \; \hat \rho =\sum_kp_k |\psi_k
 \rangle \langle \psi_k|\right].
\label{od}
\end{eqnarray} 
In the expression above, we have denoted by 
$E\left( |\psi_k \rangle \langle \psi_k|\right)$ the entanglement 
of the pure bipartite state $|\psi_k\rangle\langle \psi_k|$.
We here focus on the case of an STS  $\hat \rho_{ST}$ and recall 
that an expression for its EF could be obtained when restricting 
the optimization in Eq.\ (\ref{od}) to Gaussian pure-state decompositions only. For further convenience, we introduce the entropic function 
\begin{equation}
h(x):=\left( x+\frac{1}{2}\right) \, \ln \left( x+\frac{1}{2}\right)-\left( x-\frac{1}{2}\right) \, \ln \left( x-\frac{1}{2}\right).
\label{h}
\end{equation}
It was proven that the Gaussian EF can be expressed in terms of the
function $h(x)$ as: 
\begin{equation}
E_F( \hat \rho_{ST})=h(x_m).
\label{expr-EF}
\end{equation} 
In Eq.\ (\ref{expr-EF}), the parameter $x_m$ is given in terms of the entries of the CM \cite{PT2008}:
\begin{equation}
x_m=\frac{\left(b_1+b_2\right) \left(b_1b_2-c^2+\frac{1}{4}\right)
-2c\sqrt{{\cal D}}}{\left(b_1+b_2 \right)^2-4\, c^2}.
\label{xm}
\end{equation}
Here $${\cal D}:=\left( b_1b_2-c^2\right) ^2-\frac{1}{4}\, \left( b_1^2+b_2^2-2\, c^2\right)+\frac{1}{16}\geqq 0$$
is the main symplectic invariant. 
For any squeezed vacuum state, ${\cal D}=0,$ so that $x_m=b$, and thus 
its EF is equal to the von Neumann entropy of the reduced one-mode 
thermal state, i. e., 
$E_F\left( |{\psi}_{SV} \rangle \langle {\psi}_{SV}|\right)=h(b)$.

The difference between two classically equivalent definitions 
of the mutual information provides another measure of the total amount 
of quantum correlations in a quantum state, called discord \cite{OZ,HV}. 
Let us consider a bipartite state $\hat \rho_{AB}$ and write down 
its quantum mutual information, 
\begin{equation}
I(\hat \rho_{AB}):=S(\hat \rho_A)+S(\hat \rho_B)-S(\hat \rho_{AB}),
\end{equation}
with $S(\hat \rho)$ being the von Neumann entropy of the state $\hat \rho$. Another quantum analogue of the mutual information  is more complicated and depends on the influence on the first subsystem $A$ of the measurements 
made on the second subsystem $B$. Let us denote by $\{\hat \Pi^B_k \}$ 
a quantum measurement performed on the system $B$. The final state of the subsystem $A$ after such a measurement on the subsystem $B$ leading to the outcome $j$ is 
\begin{equation}
\hat \rho_{A|\hat \Pi^B_j}=\frac{1}{p_j}\, \mbox{Tr}_B(\hat\rho_{AB}\, 
\hat \mathbb{I}_A\otimes \hat \Pi^B_j),
\label{meas-state}
\end{equation}
In Eq.\ (\ref{meas-state}), $p_j$ is the probability of the outcome $j$:
$p_j=\mbox{Tr}(\hat \rho_{AB}\, \hat \mathbb{I}_A\otimes \hat \Pi^B_j)$. 
The quantum conditional entropy, given the non-selective measurement
$\{ \hat\Pi^B_j\}$, is a convex sum of von Neumann entropies 
of the post-measurement states\ (\ref{meas-state}) which is taken over all the possible outcomes:
\begin{equation}
S(\hat \rho_{A|\{ \hat\Pi^B_j\}})=\sum_jp_j\, S(\hat\rho_{A|\hat\Pi^B_j}).
\end{equation}
The quantum information gained about the subsystem $A$ by taking 
into account the minimal disturbance produced on it by any of all the possible measurements performed on the subsystem $B$ is the difference
\cite{OZ}
\begin{equation}
{\cal J}(\hat\rho_{AB})|_{\{\hat\Pi^B_j\}}:=S(\hat\rho_A)
-\inf_{\{\hat \Pi^B_j\} } S(\hat\rho_{A|\{ \hat \Pi^B_j\}}).
\end{equation}
The quantum $A$-discord is then defined as follows \cite{OZ}:
\begin{equation}
D_1(\hat\rho_{AB}):=I(\hat \rho_{AB})
-{\cal J}(\hat \rho_{AB})|_{\{\hat\Pi^B_j\}}\geqq 0.
\label{d1}
\end{equation}
Similarly, the quantum $B$-discord, which considers the local quantum measurements performed on the first subsystem is
\begin{equation}
D_2(\hat\rho_{AB}):=I(\hat \rho_{AB})
-{\cal J}(\hat \rho_{AB})|_{\{\hat\Pi^A_j\}}\geqq 0.
\label{d2}
\end{equation}
 
Quite recently, the above-defined discord \cite{OZ,HV} has been calculated for two-mode Gaussian states  under the approach of limiting  
the set of all one-party quantum measurements to the Gaussian 
ones \cite{PG,AD}. We were thus provided with an analytic formula 
for what is called the Gaussian discord. Moreover, according to 
Ref. \cite{PSBCL}, at least for the states analyzed here, namely, 
the STSs, the Gaussian discord is the {\em exact} discord. Thus 
the quantum discords\ (\ref{d1})  and\ (\ref{d2}) turn out to have very simple expressions in terms of one-mode von Neumann entropies:
\begin{eqnarray}
D_1^{STS}&=&h(b_2)-h(\kappa_+)-h(\kappa_-)+h(y) \nonumber\\
D_2^{STS}&=&h(b_1)-h(\kappa_+)-h(\kappa_-)+h(z). 
\label{discord}
\end{eqnarray}
Here $h$ is the entropic function\ (\ref{h}) and the symplectic eigenvalues 
$\kappa_+, \kappa_-$ are given in Eq.\ (\ref{se}).
In addition, we have used the notations:
\begin{equation}
y:=b_1-\frac{c^2}{b_2+\frac{1}{2}}, \quad z:=b_2-\frac{c^2}{b_1
+\frac{1}{2}}.
\end{equation}
Note that, for symmetric STSs ($b_1=b_2=:b$), the identity $y=z$ holds and therefore $D_1=D_2$. Moreover, for pure two-mode Gaussian states, we get 
$y=z=\frac{1}{2}$ and $D_1=D_2=h(b)$, i. e., the discord and the entanglement coincide, as expected \cite{QC2}.

\section{Evolution of a two-mode state with two local thermal reservoirs}

We consider an arbitrary two-mode field state having the annihilation operators $\hat a_1,\hat a_2,$ and the density operator $\hat \rho$. Each mode is in contact with a local thermal bath. We denote the mean photon occupancies of the two thermal reservoirs by ${\bar n}_{Rj},\, (j=1, 2)$, respectively, and the corresponding damping rates by 
$\gamma_j,\, (j=1, 2)$  . In the interaction picture, the quantum optical master equation which describes this type of coupling is
\begin{eqnarray}
&&\frac{\partial\hat \rho}{\partial t}= \frac{\gamma_1}{2}(2\hat a_1 \hat \rho \hat  a_1^{\dagger}-\hat a_1^{\dag}\hat a_1 \hat \rho -\hat \rho \hat a_1^{\dagger}\hat a_1)+\gamma_1 \bar{n}_{R1} (\hat a_1^{\dagger}\hat \rho \hat a_1+\hat a_1\hat \rho \hat a_1^{\dag}-\hat a_1^{\dagger} \hat a_1 \hat \rho -\hat \rho  \hat a_1 \hat a_1^{\dagger}) \nonumber \\
&&+ \frac{\gamma_2}{2}(2\hat a_2 \hat \rho \hat  a_2^{\dagger}-\hat a_2^{\dag}\hat a_2 \hat \rho -\hat \rho \hat a_2^{\dagger}\hat a_2)+\gamma_2 \bar{n}_{R2} (\hat a_2^{\dagger}\hat \rho \hat a_2+\hat a_2\hat \rho \hat a_2^{\dag}-\hat a_2^{\dagger} \hat a_2 \hat \rho -\hat \rho  \hat a_2 \hat a_2^{\dagger}).\nonumber\\
\label{me}
\end{eqnarray}
As in our recent work \cite{MGM} for the one-mode case, instead of 
the master equation \ (\ref{me}), we employ the equivalent differential equation for the two-mode characteristic function 
$\chi(\lambda_1,\lambda_2,t):=
\Tr \{[\hat D_1(\lambda_1) \otimes \hat D_2(\lambda_2)]\hat \rho(t)\}$. 
Here $\hat D_1(\lambda_1)$ and $\hat D_2(\lambda_2)$ are the Weyl displacement operators of the modes: 
$\hat D_j(\lambda_j):=\exp(\lambda_j\hat a^{\dagger}_j
-\lambda^{\ast}_j\hat a_j), \, (j=1, 2).$ We finally find the solution:
\begin{eqnarray} 
\chi(\lambda_1,\lambda_2,t)&=&\chi \left (\lambda_1 
e^{-\frac{1}{2}\gamma_1 t},\, \lambda_2e^{-\frac{1}{2}\gamma_2 t},0 \right) 
\, \exp \left[ -\left(\bar{n}_{R1}+\frac{1}{2}\right)\left( 1-e^{-\gamma_1t}\right) |\lambda_1|^2\right] \nonumber\\ &&\times 
\exp \left[ -\left(\bar{n}_{R2}+\frac{1}{2}\right)\left( 1-e^{-\gamma_2t}\right) |\lambda_2|^2\right].
\label{cf}
\end{eqnarray}
Let us inspect the asymptotic behaviour of the solution\ (\ref{cf}) 
of the master equation\ (\ref{me}).  When we take $t\rightarrow \infty$ 
in Eq.\ (\ref{cf}), we get the characteristic function of the two-mode thermal state imposed by the two reservoirs:
\begin{equation}
\lim_{t \to \infty}\chi(\lambda_1,\lambda_2,t)=
\exp \left[ -\left(\bar{n}_{R1}+\frac{1}{2}\right) |\lambda_1|^2 
-\left(\bar{n}_{R2}+\frac{1}{2}\right) |\lambda_2|^2\right].
\label{asympcf}
\end{equation}
Note that this two-mode steady state, which is independent of the input state, is a product state without any correlations between the modes. 

Given the structure of the time-dependent characteristic 
function\ (\ref{cf}), any input Gaussian state preserves its Gaussian
form at any time during the mode damping. In particular, an initial STS remains an STS at any subsequent time. Its evolving CM has 
the following standard-form entries:   
\begin{eqnarray} 
b_1(t)&=&b_1\, e^{-\gamma_1\, t}+\left( \bar{n}_{R1}+\frac{1}{2} \right) 
\left( 1-e^{-\gamma_1\, t} \right), \nonumber \\
b_2(t)&=&b_2\, e^{-\gamma_2\, t}+\left( \bar{n}_{R2}+\frac{1}{2} \right) 
\left( 1-e^{-\gamma_2\, t} \right), \nonumber \\
c(t)&=& c\, \exp \left[-\frac{1}{2}\left(\gamma_1+\gamma_2 \right)t \right].
\label{spt}
\end{eqnarray} 
In view of Eqs.\ (\ref{spt}), the CM of the damped STS becomes 
asymptotically diagonal:
\begin{equation}
\lim_{t \to \infty}{\cal V}(t)=\left({\bar n}_{R1}
+ \frac{1}{2} \right) \mathbb{I}_2 
\oplus \left({\bar n}_{R2}+\frac{1}{2} \right)  \mathbb{I}_2.
\label{asympCM}
\end{equation}
This means that the two-mode steady state is a product one, whose factors
are precisely the single-mode thermal states conditioned 
by the corresponding reservoirs. Thus we recover the previous general 
conclusion in the special case of an an initial STS.

To sum up, any measure of correlations in the Gaussian approach 
available for an STS, such as the entanglement of formation \cite{PT2008} 
or the quantum discord \cite{PG,AD}, can readily be applied 
for a decaying STS on account of Eqs.\ (\ref{spt}). 

\section{Evolution of a two-mode squeezed  thermal state with two identical local  thermal reservoirs}
 
What are we expecting to occur when a two-mode quantum state is subjected 
to a dissipative interaction as described by the master equation\ (\ref{me})? In general, a substantial reduction of the non-classical properties 
of the state which entails a decrease of its quantum  correlations such as entanglement and discord. More specifically, in the important particular 
case of two-mode Gaussian states, we can notice from the very beginning 
an important difference between the ways in which these two measures 
of quantum correlations actually decay. Indeed, on the one hand, 
according to condition\ (\ref{sc}), the  entanglement of the input state is expected to vanish at a finite time. This process has been called the 
{\em entanglement sudden death} in the case of qubits \cite{Yu,Damp}. 
On the other hand, it is known  that the only zero-discord two-mode 
Gaussian states are the product ones \cite{PG,AD}. Taking account 
of the time-dependent two-mode characteristic function\ (\ref{cf}), 
as well as of its steady-state form\ (\ref   {asympcf}), we infer that 
only the latter describes a product state without any correlations between the modes. Therefore, it is reasonable to believe that only asymptotically 
a damped two-mode Gaussian state could lose all its correlations, 
both quantum and classical, measured by the Gaussian discord.

For the sake of simplicity and in order to get versatile analytic results, we consider here the particular case when the two local reservoirs are identical: $\gamma_1=\gamma_2=:\gamma$ and 
$\bar{ n}_{R1}=\bar{ n}_{R2}=:\bar{ n}_R$.
In this case, the CM of an arbitrary damped two-mode Gaussian state reads: 
\begin{equation}
{\cal V}(t)={\rm e}^{-{\gamma}t}{\cal V}(0)+\left({\bar n}_{R}
+ \frac{1}{2} \right) \left( 1-{\rm e}^{-{\gamma}t} \right) \mathbb{I}_4.
\label{CM}
\end{equation} 
Here ${\cal V}(0)$ is the input CM and $\mathbb{I}_4$ denotes 
the $4\times 4$ identity matrix. Equation\ (\ref{CM}) tells us 
that an input state with no local squeezing does not change its character during damping: for instance, a symmetric state remains symmetric 
and an STS evolves as a damped STS. We restrict ourselves now to this 
latter case. When employing the entries of the time-dependent 
CM\ (\ref{CM}) in the separability condition\ (\ref{sc}), 
one finds a simple expression of the time required by a damped STS 
to reach the separability threshold:
\begin{equation}
t_s=\frac{1}{\gamma}\ln \left(1+\frac{\frac{1}{2}-\tilde \kappa_{-}}{\bar n_R}\right), \quad \left( \tilde \kappa_{-}<\frac{1}{2}\right) .
\label{ESD} 
\end{equation}
Here $\tilde \kappa_{-}$ is the smallest symplectic eigenvalue of the CM 
of the partially transposed input density matrix, which is given 
by Eq.\ (\ref{pse}). We see that in the special case of zero-temperature baths, the entanglement disappears only asymptotically. In all other cases, the quantum-classical transition occurs at finite times, i. e., 
it happens a {\em sudden death of entanglement} \cite{Yu}.
\begin{figure}[h]
\center
\includegraphics[width=5cm]{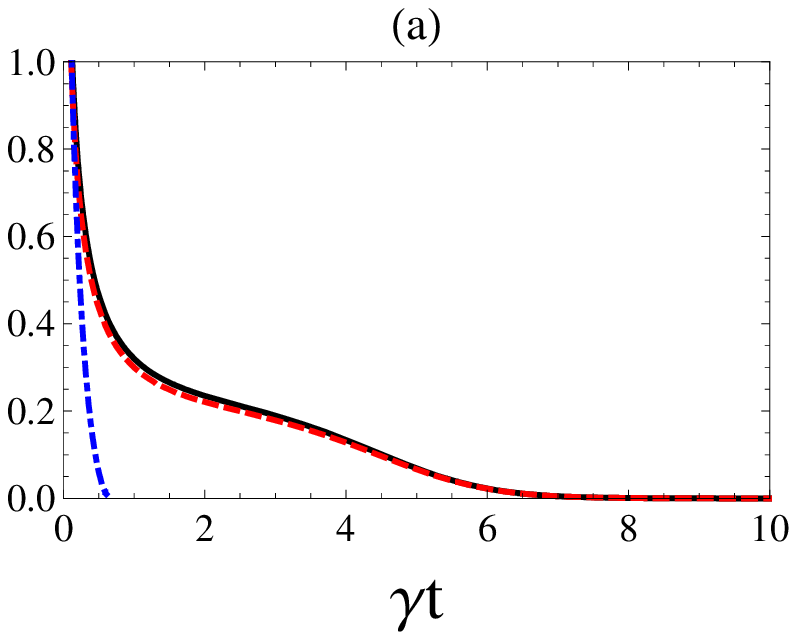}
\includegraphics[width=5cm]{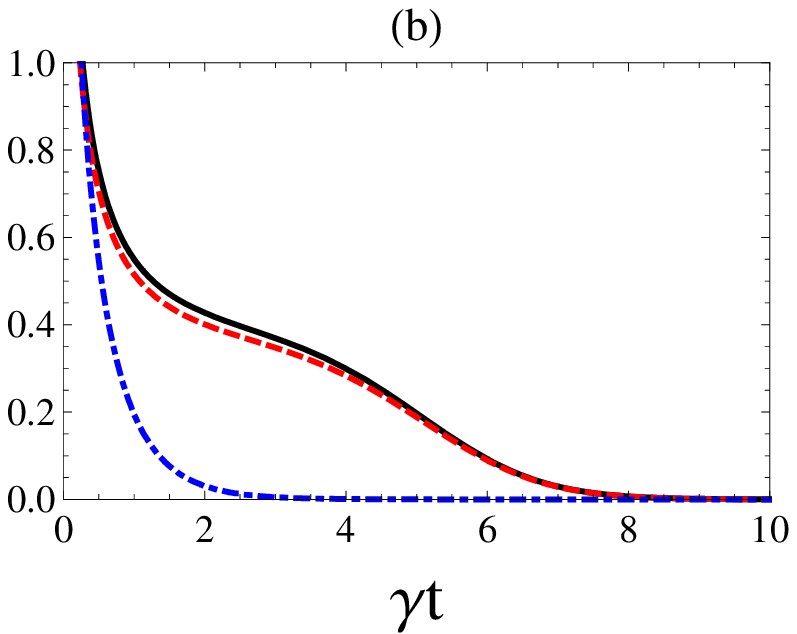}
\includegraphics[width=5cm]{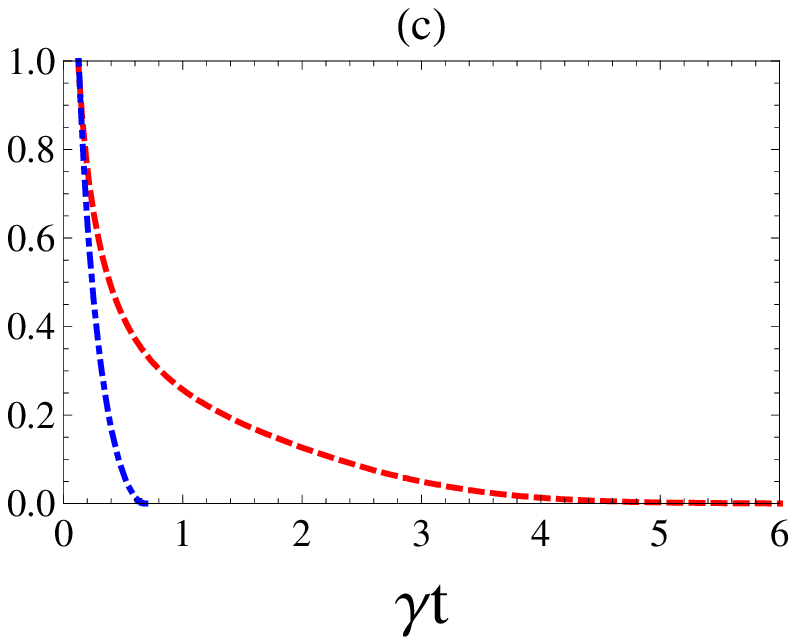}
\caption{(Color online)  Evolution of the EF (dot-dashed blue line), 
and of the discords $D_1$ (black line) and $D_2$ (dashed red line) 
for an input STS in interaction with two local identical thermal reservoirs. We have employed the following parameters. (a) The STS is characterized 
by the parameters $\bar{n}_1=10$, $\bar{n}_2=0.1$, $r=2$ and the reservoir by $\bar{n}_R=0.5$. (b) For the same input state we use $\bar{n}_R=0$. 
(c) We plot the EF and the discord $D_1=D_2$ (dashed red line) 
for an input pure state having the squeeze parameter $r=2$. The reservoir 
is noisy with $\bar{n}_R=0.5$. }
\label{fig1}
\end{figure}

In Figs. \ref{fig1}(a) and \ref{fig1}(b) we plot the evolution of the entanglement of formation, Eq.\ (\ref{expr-EF}), as well as that of the quantum discords $D_1$ and $D_2$, Eqs.\ (\ref{d1}) and\ (\ref{d2}), for an asymmetric mixed STS. The aspect of the plots follows closely our above remarks on the robustness of discord against noise in comparison with the fragility of entanglement. Note that the discords $D_1$ and $D_2$ are very close and can be distinguished only for very different values 
of the thermal mean photon occupancies $\bar n_1$ and $\bar n_2$.  
When the reservoirs are noisy, ($\bar n_R>0$), both the EF and the discords $D_1, D_2$ are strongly diminished. Contact with zero-temperature baths, 
as in Fig. \ref{fig1}(b), produces a slower decay of all correlations, 
which in this case disappear only asymptotically. In Fig. \ref{fig1}(c) 
we consider an input pure state, namely, a two-mode squeezed vacuum state 
in contact with a noisy bath. At the time $t=0$ the entanglement 
and discord coincide, but their time developments look very different. Notice that, in view of Eq.\ (\ref{CM}), a thermalized two-mode squeezed vacuum state evolves into a symmetric STS having $D_1=D_2$. 

\section{Evolution of a two-mode squeezed  thermal state with a single 
local thermal reservoir}

For finite-dimensional quantum systems it was recently found that while entanglement does not increase under local channels, other correlations 
such as discord can become larger when the input state is not pure 
\cite{bruss}. In continuous-variable settings, a similar behaviour was noticed for Gaussian discord of mixed two-mode  states under one-mode Gaussian dissipative channels \cite{Ciccarello,madsen}.
To investigate here such an interaction with analytic means and results, 
we consider an input STS having only the mode 1 in contact with a thermal reservoir. We specialize the master equation\ (\ref{me}) to the values 
$\gamma_1=:\gamma,\; \bar n_{R1}=:\bar n_{R},\; \gamma_2=0,\; \bar n_{R2}=0,$ so that the standard-form entries\ (\ref{spt}) of the damped CM become:
\begin{eqnarray} 
b_1(t)&=&b_1 {\rm e}^{-\gamma t}+\left( \bar{n}_R +\frac{1}{2}\right) \left( 1-{\rm e}^{-\gamma t}\right),\nonumber \\
b_2(t)&=&b_2,\nonumber \\
c(t)&=&c\, \exp\left(-\frac{1}{2}{\gamma t}\right). 
\label{time-dep}
\end{eqnarray} 
By insertion of the time-dependent parameters\ (\ref{time-dep}) into the separability condition\ (\ref{sc}) 
one finds  the time at which the EF of a damped STS vanishes:
\begin{equation}
t_{s}=\frac{1}{\gamma}\ln \left[1-\frac{(b_1-\frac{1}{2})(b_2-\frac{1}{2})
-c^2}{\bar n_R (b_2-\frac{1}{2})}\right], \quad
\left( b_1-\frac{1}{2}\right) \left( b_2-\frac{1}{2}\right)-c^2 <0.  
\label{ts1}
\end{equation}
We specialize Eq.\ (\ref{ts1}) to the case of a pure Gaussian input
$\left (b_1=b_2=:b,\;b^2-c^2=\frac{1}{4}\right) .$ The time of the death of entanglement\ (\ref{ts1}) is then independent of the input two-mode squeezed vacuum  state, being determined only by the field-reservoir coupling:
\begin{equation}
t_{c}=\frac{1}{\gamma}\ln \left(1+\frac{1}{\bar n_R}\right).
\label{tc}
\end{equation}
We have also checked that the time of the entanglement death 
has the same expression\ (\ref{tc}) for an input squeezed vaccuum state with additional local squeezings on both modes. Moreover, in our recent paper \cite{MGM}, we found that, for some classes of one-mode states displaying initially certain negativities of their Glauber-Sudarshan 
$P$ representation, $t_{c}$ is the ultimate time at which 
the $P$ function becomes positive due to the field interaction 
with a thermal reservoir. At the time\ (\ref{tc}) it therefore occurs 
a sudden quantum-classical transition for some types of one-mode states, as well as for any two-mode squeezed vacuum state.

As regards the evolution of the Gaussian discord, we expect it to decay
eventually very slowly and to vanish only asymptotically. Indeed,
according to Eqs.\ (\ref{time-dep}), the CM of the damped STS has 
an asymptotically diagonal form:
\begin{equation}
\lim_{t \to \infty}{\cal V}(t)=\left({\bar n}_{R}
+ \frac{1}{2} \right) \mathbb{I}_2 
\oplus b_2\, \mathbb{I}_2.
\label{asympCM1}
\end{equation}
The steady state of the field is therefore the product of two single-mode thermal states: the state of the damped mode 1, which is imposed by the thermal reservoir owing to their interaction, and that 
of the freely-evolving mode 2, which is its reduced state remaining 
constant in time and thus equal to its input at $t=0$. 

\begin{figure}
\center
\includegraphics[width=5cm]{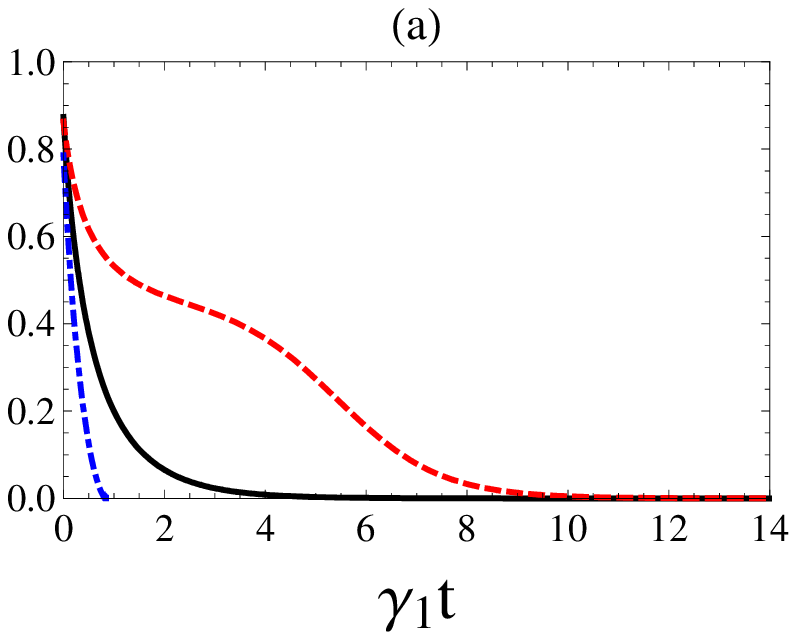}
\includegraphics[width=5cm]{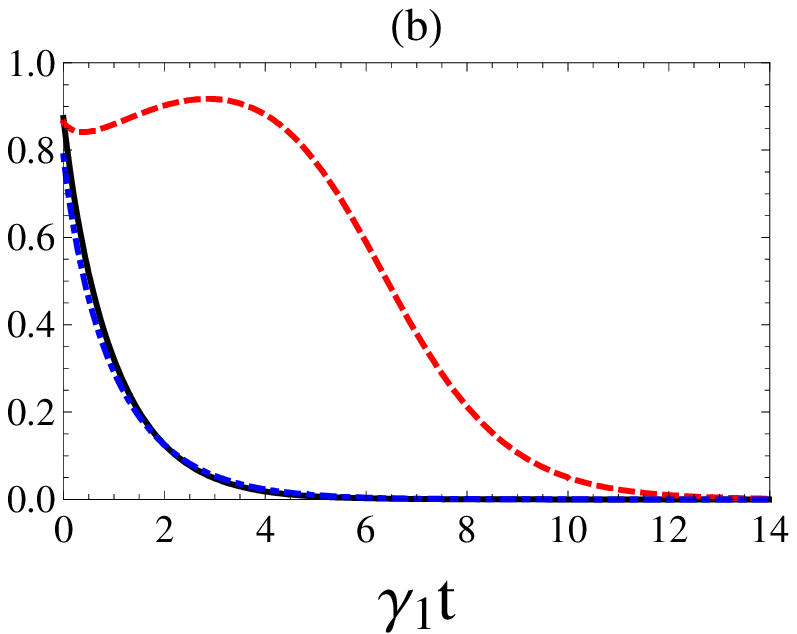}
\includegraphics[width=5cm]{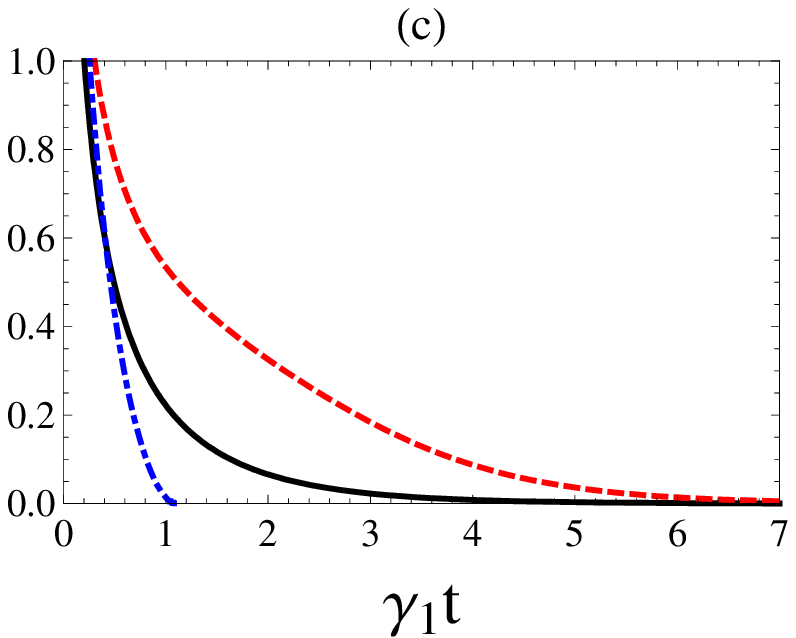}
\caption{Evolution of the EF (dot-dashed blue line), $D_1$ (black line) 
and $D_2$ (dashed red line) in a thermal bath acting on mode 1. The input state is characterized by the squeeze parameter $r=2$. The input thermal 
mean photon occupancies are $\bar{n}_1=10$, $\bar{n}_1=7$ (left and central panels) and the reservoir has (a) $\bar{n}_R=0.5$ and (b) $\bar{n}_R=0$. 
The state considered in the panel (c) is pure, while the reservoir has 
$\bar{n}_R=0.5$.}
\label{fig2}
\end{figure}

The case of a pure-state input deserves additional remarks. According 
to Eqs.\ (\ref{time-dep}), although at the moment $t=0$ the three 
measures of quantum correlations EF, $D_1$ and $D_2$ coincide, they behave subsequently quite differently because an input two-mode squeezed vacuum state evolves into an asymmetric STS. Figure \ref{fig2}(c) displays the evolution of the EF, as well as those of both discords $D_1$ and $D_2$, which are all monotonic, as predicted in Ref. \cite{bruss}. However, the discord $D_2$ corresponding to local measurements performed on the damped mode 1 survives much longer than both the EF and the discord $D_1$. 

The case of an initial mixed state can be tackled by using  
Eqs.\ (\ref{expr-EF}), (\ref{discord}), and (\ref{time-dep}) for obtaining the expressions of the EF and the discords $D_1$ and $D_2$. We plot 
in Figs. \ref{fig2}(a) and \ref{fig2}(b) their time evolution for the same input state, but with a noisy bath (a) and a zero-temperature reservoir (b). An enhancement of $D_2$ is noticed in both panels (a) and (b). The discord  $D_2$ presents a clear maximum in the latter situation and is much enhanced with respect to its value at the moment $t=0$. This can be interpreted 
as a creation of quantum correlations similar to those first explored  
for finite-dimensional systems \cite{bruss}. Moreover, in the recent 
Ref. \cite{Ciccarello} it was found that an enhancement of the discord 
$D_2$ can be noticed even when the input Gaussian state is separable. 

\section{Concluding remarks}

\begin{figure}[h]
\center
\includegraphics[width=5cm]{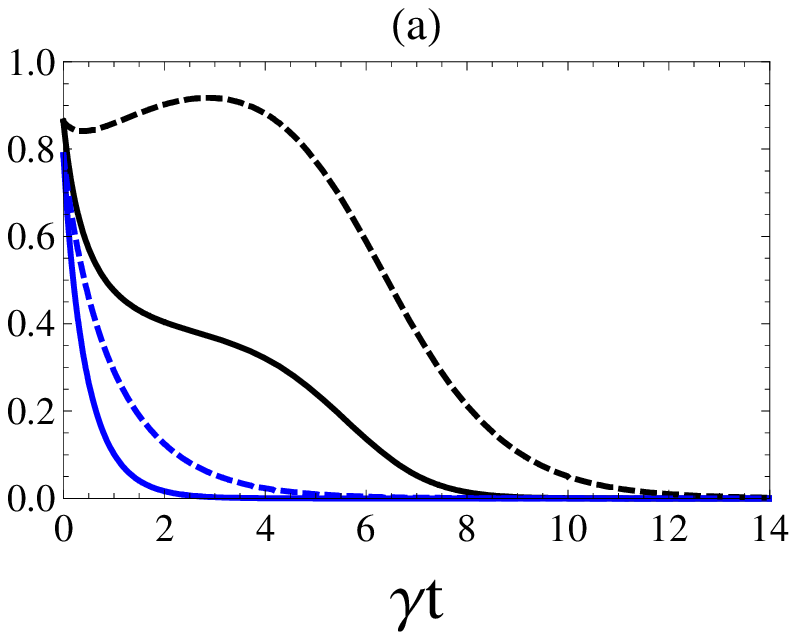}
\includegraphics[width=5cm]{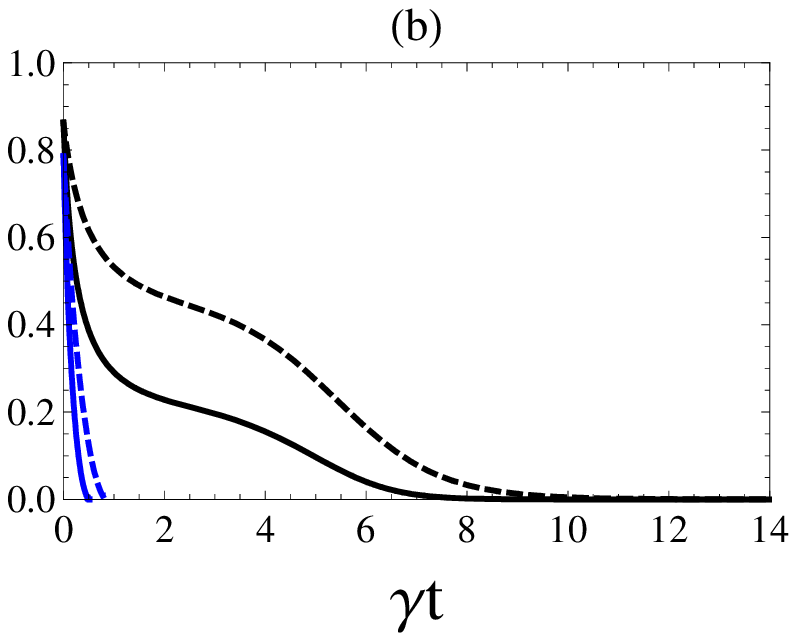}
\includegraphics[width=5cm]{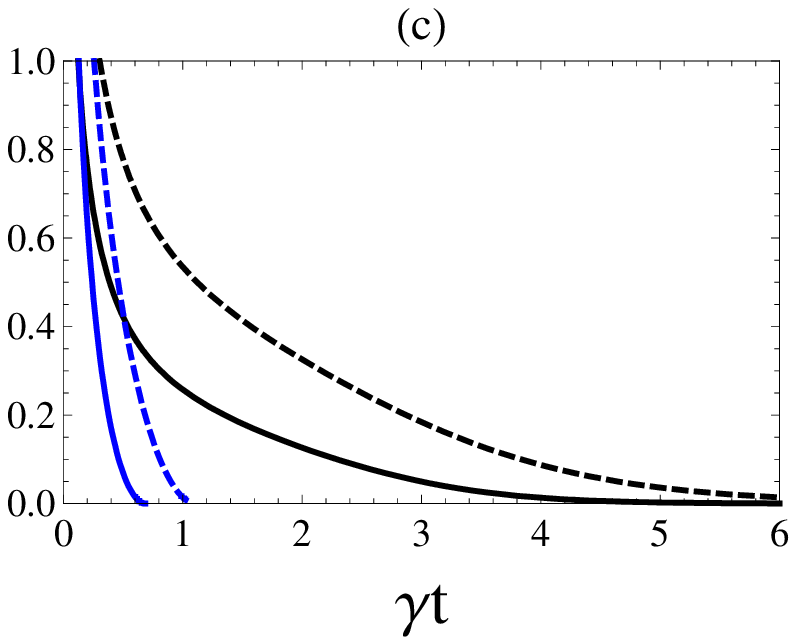}
\caption{Comparison of decays of the EF (blue curves) and of the discord 
$D_2$ (black curves) for one local bath (dashed lines) and two local identical baths (full lines) for an input state with the squeeze parameter $r=2$. The other parameters are: (a) $\bar{n}_1=10$, $\bar{n}_1=7$, 
$\bar{n}_R=0$; (b) $\bar{n}_1=10$, $\bar{n}_1=7$, $\bar{n}_R=0.5$; 
(c) an input pure state and a noisy bath with  $\bar{n}_R=0.5$.}
\label{fig3}
\end{figure}

In order to draw some conclusions on the effects produced by local  dissipation on quantum correlations of an STS, we compare now the decay of entanglement and discord for the two situations studied above.
 Figure \ref{fig3} displays our results for the EF (blue curves) and $D_2$ (black plots) in the cases of both one and two local identical reservoirs for a mixed STS (panels (a) and (b)) and for a pure Gaussian state (c). 
We represent here the case of zero-temperature reservoirs (panels (a) and (c)) to show a better preservation of all correlations in comparison 
with the noisy bath considered in panel (b).  We can see that in all cases  both Gaussian discords $D_1$ and $D_2$ survive longer than the EF. This is expected because only asymptotically the damped Gaussian state becomes a product one. Thus the Gaussian discord, which measures the whole amount of quantum and classical correlations, proves to be quite robust against dissipation in all the above-mentioned situations.
However, only for a configuration with one local thermal bath, there is 
an enhancement of the discord $D_2$. Since this can be larger than the discord of the input state, it means that the field-reservoir interaction generates quantum and clasical correlations of the discord type.
 A final conclusion arising from Fig. \ref{fig3} is quite interesting. 
In all the analyzed situations (mixed or pure input states, noisy 
or zero-temperature reservoirs), the configuration with one local thermal bath performs better than that with two local identical baths. 
This is valid when analyzing  both the magnitude of correlations 
and their preservation in time.

\ack{This work was supported by the Romanian National Authority for Scientific Research through Grant PN-II-ID-PCE-2011-3-1012 for the University of Bucharest.}

\end{document}